\documentclass[11pt,a4paper]{article}
\usepackage{jheppub}
\usepackage{enumerate}
\usepackage{amsmath}

\usepackage[hyperref,svgnames]{xcolor}

\usepackage{subfigure}
\usepackage{floatrow} 

\newcommand{\gsim}{\lower.7ex\hbox{$\;\stackrel{\textstyle>}{\sim}\;$}}
\newcommand{\lsim}{\lower.7ex\hbox{$\;\stackrel{\textstyle<}{\sim}\;$}}

\newcommand{\thb}{{\bar\theta}}

\newcommand{\MeV}{{\, {\rm MeV}}}
\newcommand{\GeV}{{\, {\rm GeV}}}
\newcommand{\TeV}{{\, {\rm TeV}}}

\def\beq{\begin{equation}}
\def\eeq{\end{equation}}
\def\bea{\begin{eqnarray}}
\def\eea{\end{eqnarray}}
\def\bitem{\begin{itemize}}
\def\eitem{\end{itemize}}
\newcommand{\bec}{\begin{center}}
\newcommand{\eec}{\end{center}}
\newcommand{\ba}{\begin{array}}
\newcommand{\ea}{\end{array}}

\DeclareMathOperator{\Tr}{Tr}

\def\bar#1{\overline{#1}}

\def\inv{^{\raise.15ex\hbox{${\scriptscriptstyle -}$}\kern-.05em 1}}
\def\lbar{{\lower.35ex\hbox{$\mathchar'26$}\mkern-10mu\lambda}} 

\let\<=\langle
\let\>=\rangle

\let\+=\uparrow

\let\Si=\Sigma

\def\OO{\mathcal{O}}

\newcommand{\mpl}{M_{\rm pl}}

\begin{document}

\title{QCD, Flavor, and the de Sitter Swampland}
\author[]{John March-Russell,}
\author[]{Rudin Petrossian-Byrne}  
\affiliation[]{Rudolf Peierls Centre for Theoretical Physics, University of Oxford,
Oxford, OX1 3PU, UK}
\emailAdd{John.March-Russell@physics.ox.ac.uk}
\emailAdd{rudin.petrossian-byrne@physics.ox.ac.uk}

\abstract{The refined swampland de Sitter conjecture (SdSC) is a proposed constraint on the form of the total potential in a theory
including quantum gravity.  According to this conjecture potentials possessing metastable de Sitter vacua are in the swampland of effective field
theories that cannot descend from a theory with gravity.  It is known that in the Standard Model (SM), as the quark masses and ${\bar\theta}$-parameter
are varied, IR-calculable metastable states in QCD appear (for $N >2$ light quarks) and we discuss in detail their properties. 
We argue that it is possible that the SdSC excludes the values of quark masses and ${\bar\theta}$ for which these metastable states can arise,
leading to a possible surprising connection between quantum gravity and aspects of low-energy flavor phenomenology.
The \emph{observed} values of the quark masses and QCD ${\bar\theta}$-parameter are consistent
with the SdSC.
If, in addition, as partially indicated by
large-$N_c$ and semi-classical analysis, pure $SU(3)$ Yang-Mills theory has metastable states at ${\bar\theta}=0$
(this to our knowledge is not definitively known) then much of the a-priori SM parameter
space might be eliminated.  In particular, if quark Yukawa couplings are kept fixed, the limit of large electroweak vacuum expectation
$v_{EW}\gsim 50\TeV$ could be excluded by the SdSC, possibly shedding a new light on the hierarchy problem. 
We argue that these statements are robust against the addition of a quintessence field unless extreme fine-tuning is allowed.
}

\maketitle 
\section{\label{intro}Introduction}

There is mounting evidence that not all consistent low-energy quantum effective field theories (EFTs) can descend from
a theory including quantum gravity.  Those EFTs that cannot so descend are said to belong to the `swampland' while those that can
are in the `landscape' of consistent theories including gravity.   The swampland program \cite{Vafa:2005ui,Brennan:2017rbf} then has the aim
of providing sharp criteria which delineate the swampland from the landscape.  At present
most of the suggested criteria take the form of conjectures, often with unknown $\OO(1)$ constants appearing, and having varying
levels of support.   The most established results include the statement 
that exact continuous (and likely discrete) global symmetries satisfying certain locality properties are incompatible with quantum 
gravity\cite{Banks:1988yz,Kamionkowski:1992mf, Holman:1992us, Kallosh:1995hi,Argyres:1998qn,Banks:2010zn,Harlow:2018tng,Fichet:2019ugl,Daus:2020vtf}
leading to a Swampland Global Symmetry Conjecture (SGSC), and the closely connected Weak Gravity Conjecture concerning the limit
of small gauge coupling \cite{ArkaniHamed:2006dz}.  

One of the most striking of the conjectured constraints is the (refined) swampland de~Sitter conjecture (SdSC)
of \cite{Obied:2018sgi,Ooguri:2018wrx} which states, in the 4-dimensional case, 
that the total low-energy potential $V(\{\phi_i\})$ must either satisfy
\bea
| \nabla V |  & \geq & c \frac{V}{\mpl} ~~\label{eq:rdSc1} \\
{\rm or}\qquad {\rm min}( \nabla_i \nabla_j V ) & \leq & - c'  \frac{V}{\mpl^2} ~~\label{eq:rdSc2}
\eea
for $\OO(1)$ coefficients $c,c'>0$,  and where $\nabla$ denotes a derivative with respect to all scalar field directions $\{\phi_i\}$.
Thus according to this conjecture \emph{potentials possessing metastable de~Sitter vacua are in the swampland, as are potentials with
regions of field space that are too `flat' if} $V>0$.  

So far the dominant use of this conjecture has
been in the cosmological context, for example the analysis of Ref.~\cite{Agrawal:2018own}. 
It has been argued that a possible early epoch of cosmological inflation
is severely constrained, and that the apparent presently-observed
cosmological acceleration must be due to a time-evolving quintessence field, $\varphi(t)$, and not a true cosmological constant.  The
validity of the SdSC is controversial, with arguments being made both for and against.  It is also possible that a further refinement
of the conjecture is necessary beyond fixing the presently unknown, but believed to be $\OO(1)$, constants $c,c'$.  For our
purposes we \emph{assume that the SdSC is correct as stated} and explore its possible ramifications for low-energy physics.

Specifically we make the surprising claim that it is possible that the SdSC limits the allowed values of the quark masses and
QCD-${\thb}$ parameter.   We find that the \emph{observed} values of the quark masses and ${\thb}$-parameter are consistent
with the SdSC, but variations away from the observed values lead to IR-calculable potentials
which, as functions of the light pseudo-Nambu-Goldstone boson (pNGB) fields, $\pi^a(x)$, possess metastable
de Sitter vacua.  The difference, $\Delta V$, in the value of the potential energy density between the true ground state 
and these metastable states satisfies, $H_0^4\ll \Delta V\ll \mpl^4$, where $H_0$ is the present value of the Hubble parameter.
In addition these metastable vacua occur for field values $\langle \pi^a(x) \rangle \lsim f_\pi \ll \mpl$.  Because of this large parametric separation 
in scales, our effective field theory analysis of the metastable state structure is under good control.   Moreover, we are here making what we believe
to be the reasonable statement that if the SdSC makes sense in its current form then it should apply not only to `fundamental' scalar fields and their
potentials but also composite scalar fields and their potentials arising from strong-coupling dynamics.  (If this were not the case then a straightforward
extension of our results imply that it would be easy to construct simple strong-coupling hidden sector models that would lead to a very-long-lived
de Sitter phase, and thus in practical terms invalidate the constraints on early and late cosmology from the SdSC.)

Thus we make the claim that regions
of quark-mass-parameter and ${\thb}$-parameter space might be forbidden by consistency with quantum gravity!
Indeed, a connection between quantum gravity consistency and the detailed properties of QCD and the flavor
sector of the SM is in fact \emph{already implied} by other better established (though we emphasise not yet proven) swampland conjectures.
For example, the Swampland Global Symmetry Conjecture already states that, for the pure SM, it is inconsistent with quantum gravity 
for any two equally-charged quarks to have Yukawa couplings that simultaneously vanish or be exactly equal.  

Although in its own right this statement regarding the violation of global symmetries
is a fascinating connection between quantum gravity and low-energy physics, practically speaking this only excludes only a vanishingly
small set of the a-priori SM parameter space. (Naively of measure zero, but in fact likely a thicker
set of size $\exp(- \mpl^2/ \Lambda^2)$ where here $\Lambda$ is a suitable cutoff of the low-energy theory related to the tension of strings
\cite{Daus:2020vtf,Hebecker:2017uix}.)  On the other hand \emph{we will argue that the refined swampland de Sitter conjecture plausibly excludes an} $\OO(1)$
\emph{subset of the a-priori allowed SM parameter space, and thus is potentially a much more powerful restriction on the low-energy features
of the SM.}

\subsection{The primary idea}

As has been known for some time \cite{Witten:1980sp,Creutz:1995wf,Smilga:1998dh} and as we will argue in detail in section \ref{ChiralLagmetastable}, QCD can exhibit for $N>2$ light quarks both a true ground state \emph{and metastable states} as a function of the light
pseudo-Nambu-Goldstone boson fields, $\pi^a$ (here $a=1,\ldots,N^2-1$).

\begin{figure}
\centering
  \includegraphics[scale=0.2]{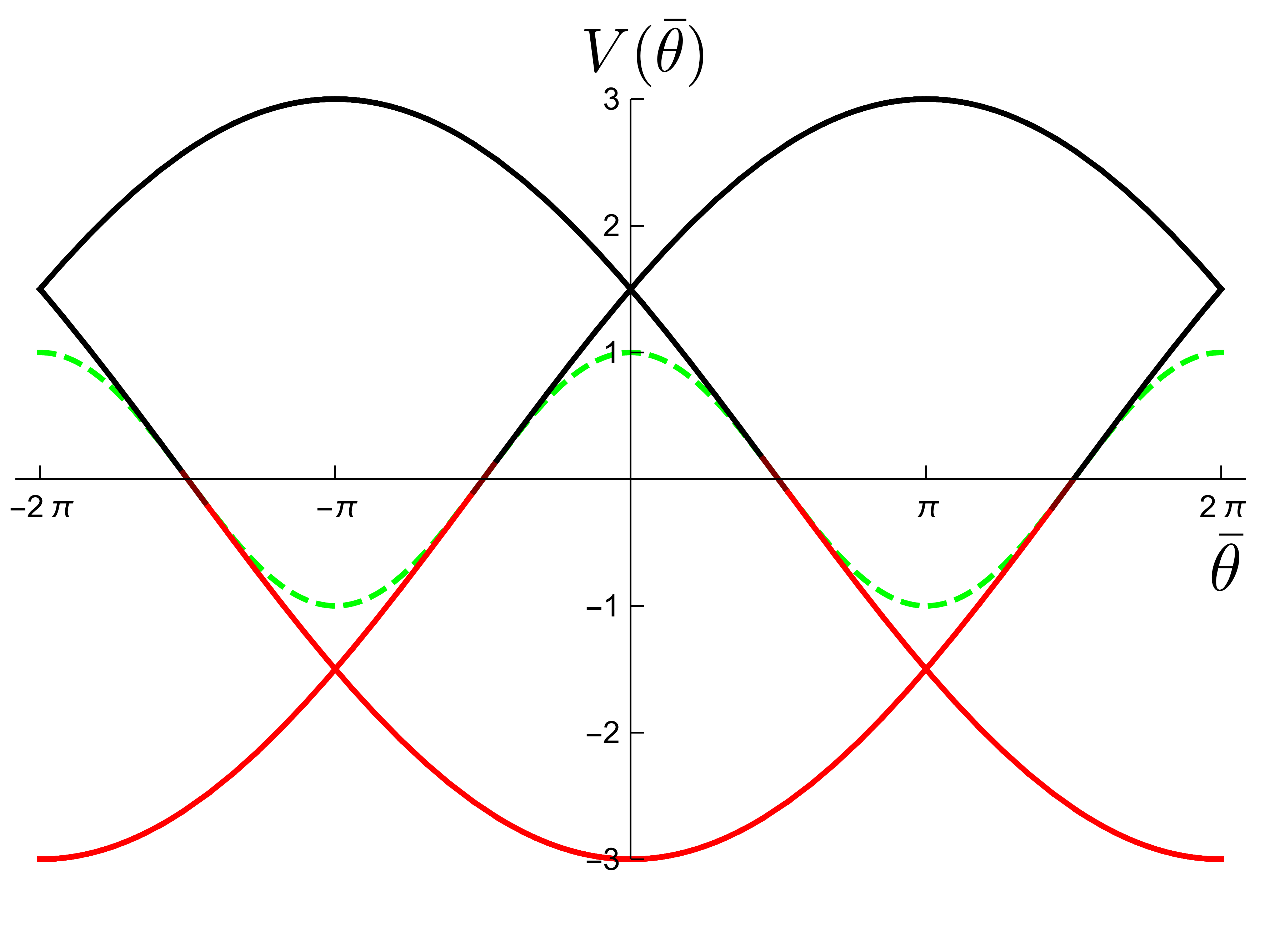}
\caption{The behavior, as a function of $\thb$, of the potential energy density of the critical points of the three-light-flavor leading-order chiral Lagrangian
when quark masses, $m$, are equal \cite{Creutz:1995wf,Smilga:1998dh}.  Red sections of curves are local minima, dashed green are saddle points,
and black sections local maxima.  The true ground state, the lowest segment of the red curves, is a $2\pi$-periodic, non-analytic function of $\thb$.  In the range
$\pi/2 < \thb < 3\pi/2$ there are two local minima, \emph{with the upper section of the red curve being a metastable state split from the ground state by}
$\OO(m\Lambda_{qcd}^3)$.  At $\thb =\pi$ the Dashen phenomena \cite{Dashen:1970et} occurs, CP is spontaneously broken, and there are two
degenerate ground states.
\label{fig:3equalmasses}} 
\end{figure}

\begin{figure}
\centering
  \includegraphics[scale=0.8]{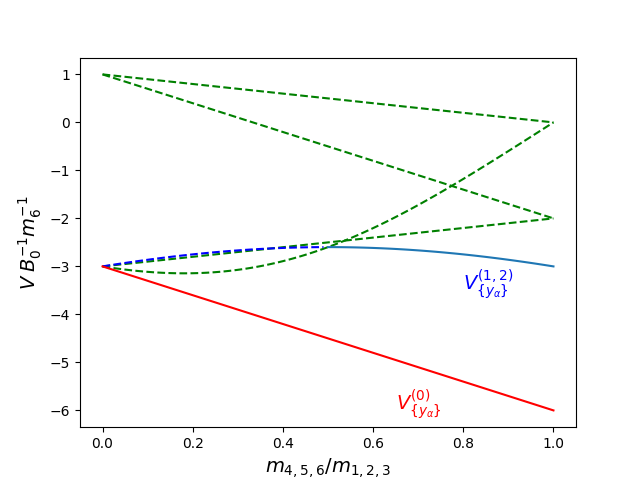}
\caption{An example of metastable states appearing in the vacuum structure of QCD at $\thb=0$ as quark masses are varied. Here we divide 6 light quarks into two sets of three equal mass quarks, $m_1=m_2=m_3$ and $m_4=m_5=m_6$, and continuously vary the mass ratio $0\leq m_6/m_1 \leq 1$.  $V^{(0)}_{\{y_\alpha\}}$ in red corresponds to the global minimum `branch', while
$V^{(1,2)}_{\{y_\alpha\}}$ in blue are a pair of degenerate branches which turn from saddle points (dashed line) to metastable states (solid line). Some, but not nearly all, other branches of saddle points are also shown. $B_0 \sim \Lambda_{QCD}^3$ is a parameter of the chiral Lagrangian related to the pion mass. The typical difference in energy densities between the ground state and the metastable branches is $\sim {\rm few}\times B_0 m_6 \sim  {\rm few}\times\Lambda_{qcd}^3 m_6$. \label{fig:Vbranches}} 
\end{figure}

Suppose the QCD- and flavor-sector parameters of the SM which here we collectively call $\{ y_\alpha\}$, which include both heavy and light quark masses and the QCD
${\thb}$-parameter, are
allowed to vary from their observed values.\footnote{For the purposes of this paper we will only consider the
quark masses and ${\thb}$-angle as variable parameters, and fix all other SM parameters such as the electromagnetic fine structure constant and the QCD
scale $\Lambda_{qcd}$.  
Other `hidden' parameters are possible too, eg, the scale of a spontaneous breaking $SU(N_c)\rightarrow SU(3)_c$ if we want to smoothly extend
consideration to $SU(N_c)$ theories of the QCD-gauge group.}  
Critical points of the QCD-sector potential will move continuously as a function of the $\{ y_\alpha\}$ parameters and may change their character, between local minimum, saddle point and local maximum.
Let the energy density of these discrete `branches', labelled by $n=0,1,\ldots$, be $V^{(n)}_{\{ y_\alpha\} }(\pi^a)$. 
We choose the convention
that at any given value, $\{ y_\alpha\}$, of the SM parameters the true ground state branch is $n=0$ while $n=1,\ldots $ label the 
potential functions for the various non-ground-state branches in ascending order of energy density. Note, importantly, that as the
SM parameters $\{ y_\alpha\}$ vary the number of metastable branches can change. Branches can also cross or merge. Thus in general the behavior of the ground state and excited
branches is a non-analytic and extravagant function of the parameters $\{ y_\alpha\}$.  We illustrate this behavior in figure~\ref{fig:3equalmasses} for the simple case
of three equal mass light quarks as a function of $\thb$, and in figure~\ref{fig:Vbranches} for six light quarks divided into two groups of three equal
mass quarks, as a function of the mass ratio, and for $\thb=0$.\footnote{Although, strictly speaking, three quarks of exactly equal mass is a point in parameter space
forbidden by the SGSC, it is  important to note that the form of the curves is an analytic function of the $\{ y_\alpha\}$, so if we move very slightly away from exact equality of masses, or special points like $\thb=0$, the number and properties of the critical points of the potential is almost everywhere unchanged, with the exceptions being points where curves cross. So the situation illustrated in figures~\ref{fig:3equalmasses} and \ref{fig:Vbranches} is a good guide as we explicate in detail in section~\ref{sec:thePotential}.}

Most importantly, and as is well known, using the power of chiral Lagrangian techniques the $V^{(n)}_{\{ y_\alpha\} }(\pi^a)$ are
\emph{IR-computable} functions of the $\pi^a$'s not depending on details of the UV completion. 
So as to compare hypothetical worlds on a like-for-like basis, we demand that in the ground state of the SM, so in the branch $V^{(0)}_{\{ y_\alpha\} }(\pi^a)$, the total vacuum energy is (close to -- we quantify this in section \ref{quint}) zero for each and every choice of $\{ y_\alpha\}$.  Minimally we do this by tuning a necessarily
$\{ y_\alpha\}$-dependent additive constant in the potential, which may or may not be, for example, the result of some continuous or discrete neutralization or relaxation
mechanism \cite{Abbott:1984qf,Brown:1988kg,Bousso:2000xa,Feng:2000if,Graham:2019bfu} as long as it itself is not in conflict with the SdSC and other
swampland constraints.  We of course do not currently have an accepted good theory of this tuning, but this tuning cannot simultaneously set both
the effective vacuum energy density of the metastable state(s) and the stable ground state to zero.  Our working assumption is that the SM ground state is the
state that must have vacuum energy close to zero. 
Then, if the SM plus gravity were a complete description of low-energy, $E\lsim \Lambda_{qcd}$, 
physics, with no ultra-light feebly-coupled fields present, and given the stated assumptions, the SdSC
would immediately forbid those values, $\{ y_\alpha^*\}$, of the SM parameters for which metastable states exist. 

Although the vacuum structure of the strong sector is determined by the $N$ `light' quarks, $m_q \ll 4\pi f_\pi \sim  4 \pi \Lambda_{qcd}$, the value of $\Lambda_{qcd}$
itself, however, is of course sensitive to the UV initial condition on $g_{3}$ as well as, to a lesser extent, the masses of the `heavy', $M_q \gtrsim 4 \pi \Lambda_{qcd}$, quarks, which stop contributing to the running at scales $\mu<M_i$.  Concerning this, if
$\Lambda_{(n_f)}(g_{3,\rm UV})$ is the value of the strong scale computed with $n_f$ quarks always active, 
the effect of a heavy quark mass threshold is, to leading order \cite{Deur:2016tte},
\begin{equation}
	\Lambda_{(n_f-1)} = \Lambda_{(n_f)}^{(33-2n_f)/(35-2n_f)} M_{1}^{2/(35-2n_f)}~
\end{equation}
where $M_1$ is the mass of the heaviest quark.
So, as the heavy quark masses are lowered, the value of $\Lambda_{qcd}$ reduces, and in principle more quarks could become light as some of the heavy quark masses are reduced, or vice versa.  However this effect is quite weak. 
In this paper we limit ourselves to simply considering `QCD with $N$ light quarks' in its generality, for different mass ratios and vacuum angles, irrespective of
what $\Lambda_{qcd}(g_{3,\rm UV})$ and the $M_i$ are. In reality, it is of course to be expected that these parameters and others are set by UV quantum gravity dynamics
and it is certainly possible that one may not freely vary each individually.  In fact, if the SdSC were true, our results make the surprising point that the correlated
dependence of the low-energy parameters on the underlying UV parameters must be such so that the values of $\{ y_\alpha\}$ we find to be in the swamp could never be attained.

In section \ref{ChiralLagmetastable}
we delineate, in the case where there are two or more light quarks, $N\geq 2$, the regions of quark mass and ${\thb}$-parameter space which
are excluded by this criterion.  The region of metastable states is determined by a function of the light-quark mass ratios and $|{\thb}|$.  For example, in
the limit that the two lightest masses become degenerate $m_{N-1} \rightarrow m_N$ (our convention is that $m_N$ is the mass of the lightest quark)
our results are captured by the following single condition
\begin{equation}
\label{eq:master_eq}{}
	\sum_{i=1}^{N-2} \sin^{-1}\left( \frac{m_N}{m_{i}}\right) > \pi - |\thb|, \quad \quad \thb \in (-\pi,\pi ]~.
\end{equation}
If satisfied, the theory will feature a metastable state as derived from the leading order chiral Lagrangian (if the inequality
is saturated a higher-order analysis is necessary).
The general condition, valid for arbitrary light quark mass ratios, is made precise in section \ref{sec:thePotential}.

As the number of light quarks increases, larger and larger regions of the parameter space possess metastable states and so are excluded.   For instance,
in the case of all light quark masses equal, we find $N > 4$ possesses metastable states for $\thb=0$ ($N=4$ at $\thb=0$ requires a higher order analysis that we will cover in a later work).  This continues to hold for a range of mass ratios away from exact equality.  For example in figure~\ref{fig:Vbranches} we show how a branch goes from being a saddle point to a metastable state as a particular combination of masses is changed.
As $\thb$ becomes non-zero the range of ratios $\{m_\alpha/m_\beta\}$ in the swamp increases. 
The analysis shows that for $|\thb|>\pi/2$ even for $N=3$, close relatives of our world have metastable states.
For the theory to be `safe' from the swamp for all $\thb$ one requires\footnote{The generalisation to N quarks is straightforward, see eq.(\ref{eq:safe_zone}).}
\begin{equation}\label{eq:N3safe}
	\frac{1}{m_3} > \frac{1}{m_2}+ \frac{1}{m_1}~,
\end{equation}
a condition curiously satisfied by the SM in the up, down and strange quark masses (see figure~\ref{fig:3lqTheta_arrow}).
As well as providing analytic expressions, we develop a diagrammatic method (a `fan diagram') for identifying metastable states, as well as
more general critical points.

This analysis summarized above assumes the IR theory is the SM plus gravity, so we have not yet considered the
effects of additional ultra-light feebly-coupled fields which may be motivated for two reasons.  First, 
cosmological observations appear to demand that we are presently in an epoch of accelerated cosmological
expansion.
As mentioned, a net positive cosmological \emph{constant} term is in contradiction with the SdSC,\footnote{Given the apparent
immense theoretical difficulty and fine-tuning implied by such a tiny positive
cosmological term, we do not consider it \emph{definitely} established that the observations are explained by an
effective non-zero vacuum energy rather than, say, an apparently highly unusual but not so extremely-fine-tuned alternative
cosmology without a positive vacuum energy.  For the purposes of this paper we assume that the observational claims of acceleration are
correct and, moreover, there \emph{is} at present a tiny positive effective vacuum energy density.}
and therefore the approach taken in the literature has been to assume that there exists an ultra-light quintessence field, $\varphi(t)$, which is evolving
in a potential with suitable tiny effective vacuum energy density without a local minimum and which (marginally) satisfies the SdSC \cite{Agrawal:2018own}.  
If this is correct then we must include the quintessence field in our analysis too.\footnote{Here we are ignoring the stimulating ``Thermal Dark Energy" proposal of \cite{Hardy:2019apu} which avoids quintessence fields.}
In section \ref{quint} we address this modification and argue that, 
unless an extreme fine-tuning is allowed, it is impossible to include a quintessence field in a manner consistent with the SdSC that does not also lead to a collapse into an AdS state on excessivley short timescales.
Thus the SdSC still excludes the regions of parameter space with QCD metastable states. 

Second, we have excluded the possibility of a light QCD axion relaxing the ${\thb}$ angle to zero \cite{Peccei:1977hh,Weinberg:1977ma,Wilczek:1977pj}. 
We view this as less of an issue as there are 
non-Peccei-Quinn-Weinberg-Wilczek solutions to the strong-CP problem that do not involve adding new states in
the deep IR of the SM \cite{Nelson:1983zb,Barr:1984qx,Babu:1989rb,Barr:1991qx,Babu:2001se}. Nevertheless, it is interesting to consider the effect of the
axion solution to the strong-CP problem on our reasoning, and we will address this in a companion paper.

In section~\ref{sec:N=0} we turn to the question of whether the case of no light quarks, $N=0$, namely, pure $SU(3)_c$ Yang-Mills theory,
possesses metastable states and thus is in the swampland.  There are of course two ways this situation could occur: Either all the quark Yukawa couplings
could be $\gg 0.01$ with the electroweak vacuum expectation value (vev)
fixed.  Or, most interestingly, the Yukawa couplings can retain their standard values while the electroweak vev is taken to be $\gg 50 \TeV$.  In this second
case our thinking then possibly leads to a new perspective on the hierarchy problem.   Fascinatingly, as has been known for a long time, large-$N_c$ analysis
of the pure $SU(N_c)$ theory strongly implies an $\OO(N_c)$ number of metastable states (we review this statement
and its refinements in section~\ref{sec:N=0}), though this analysis breaks down for small $N_c$.
Moreover other, related, semi-classical arguments possibly indicate that even $SU(3)$ Yang-Mills theory might have metastable states at ${{\bar\theta}}=0$.
To our knowledge there are no lattice studies of this issue, so it is not definitively known whether pure $SU(3)$ has metastable states or not.\footnote{We thank
Mike Teper for discussions of this issue.}  If we assume
that  it does,  much of the a-priori SM parameter space is eliminated by the SdSC, in particular the limit of large electroweak vev
$v_{EW}\gsim 50\TeV$ is excluded (if quark Yukawa couplings are kept fixed).  
We again argue that these statements are robust against the addition of a quintessence field unless extreme fine-tuning is allowed.
Thus it is possible that the SdSC sheds a significant new light on the hierarchy problem.   In a companion paper we discuss the limit of
large positive Higgs $m_H^2$ parameter in light of the Swampland Program.  

 We now turn to the details of our analysis.

\section{\label{ChiralLagmetastable}Metastable states of $SU(3)_c$ with $N\geq 2$ light quarks}
\label{sec:thePotential}

We here focus on the study of the metastable states of the color and quark sector of the SM for
two or more light quarks (but not so many that UV asymptotic freedom and IR
confinement and $SU(N)_L \times SU(N)_R \rightarrow SU(N)_V$ chiral symmetry breaking are lost).  In this case we may use the power of chiral Lagrangian
techniques to investigate the vacuum and metastable state structure of theory as a function of the light quark masses
and the QCD ${\thb}$-angle \cite{Witten:1980sp,Creutz:1995wf,Smilga:1998dh,Dubovsky:2010je}.   Written in terms of the Nambu-Goldstone field $\Si(x) = \exp(2i\pi^a(x) T^a/f_\pi)\in SU(N)$, the relevant terms in the chiral
Lagrangian are  
\bea\label{chiralL}
{\cal L} = &&\frac{f_\pi^2}{4} \Tr\left( \partial_\mu \Si^\dagger \partial^\mu \Si \right) -B_0 \Tr\left( e^{-i\thb/N} M_q^\dagger  \Si +e^{i\thb/N} \Si^\dagger M_q \right)~,
\eea
where $M_q$ is the $N \times N$ quark mass matrix, and $B_0>0$ is a mass-dimension three parameter, $\OO(\Lambda_{qcd}^3)$, whose precise
value can be related to the pion mass.
Using field redefinitions it always possible to take, without loss of generality, the quark mass matrix to be diagonal and
real, $M_q= {\rm Diag}(m_1,m_2,\ldots,m_{N})$, with $m_1\geq m_2\geq \ldots \geq m_{N}$.

\subsection{Critical points of the potential}\label{sec:criticalpoints}

As shown in, eg, \cite{Dubovsky:2010je} all spacetime-independent extrema of the potential eq.(\ref{chiralL}) can be written in diagonal form 
\begin{equation}
\label{eq:Vacuum_ansatz}
	\Si(x) = e^{i\thb/N}{\rm Diag}(e^{\phi_1},e^{\phi_2},\ldots, e^{\phi_{N}})~.
\end{equation}
Here an overall $\exp(i\thb/N)$ has been factored out for convenience.  Special unitarity requires
\beq
\label{eq:suN_constraint}
\phi_1 + \dots + \phi_N + \thb = 0 \quad \text{mod} \; 2 \pi~.
\eeq
Subject to this constraint $(\phi_i + \thb/N)f_\pi \equiv \langle \pi^i \rangle$ defines a useful linear-recombination of the vacuum expectation values
of the neutral ($SU(N)$ Cartan sub-algebra) pNGB fields.  The relevant potential is then 
\begin{equation}\label{eq:V(phi)}
  	V(\phi_i) = - B_0  \sum_i^N m_i \cos\phi_i~.
\end{equation}
Each angle is associated to a quark mass.
Differentiating eq.(\ref{eq:V(phi)}) subject to eq.(\ref{eq:suN_constraint}) gives a condition for a critical point in terms of the tower of identities:
\beq
\label{eq:sin_idts}
\sin\phi_1 = \frac{m_2}{m_1} \sin\phi_2 = \dots  =  \frac{m_N}{m_1} \sin\phi_N~.
\eeq
We see therefore that the angles $\phi_i$ must be spread out like an ordered fan, as shown in figure~\ref{fig:critical_points}.  


Taking $\phi_{i<N}$ as independent, the relevant Hessian determining the nature of the critical points is
\begin{equation}
	H_{ij} = \delta_{ij} m_i \cos\phi_i + m_N \cos\phi_N~,
\end{equation}
from which one sees that a necessary condition for positive-definiteness is $\cos{\phi_{i<N}}>0$ as in figure~\ref{fig:local_minima}. If we also
have $\cos\phi_N > 0$, \emph{the critical point is guaranteed to be a local minimum}.  The more general condition is given by\footnote{One way to see this is by studying the characteristic polynomial of the symmetric matrix $H_{ij}$. The condition eq.(\ref{eq:pos_def}) can be seen to ensure that there are no negative roots (i.e. eigenvalues) by Descartes' rule of signs.}
\begin{equation}
\label{eq:pos_def}
	{\rm det}\left( H_{ij}\right) = \sum_i^{N} \prod_{j\neq i}^N m_j \cos{\phi_j} > 0~,
\end{equation}
\begin{equation}
\label{eq:pos_def_function}
	\implies C(\phi_N)=  1 + \frac{m_N\cos\phi_N}{m_1 \cos\phi_1} + \dots + \frac{m_N \cos\phi_N}{m_{N-1} \cos\phi_{N-1}}>0~.
\end{equation}
$C$ is chosen to depend on $\phi_N$ as all other $\phi_i$ at a local minimum point unambiguously follow from eq.(\ref{eq:sin_idts}).

\subsection{Metastable states at equal quark masses}\label{sec:equalmasses}
A particularly simple case in which analytic expressions are straightforward to derive occurs when all masses are equal, $m_i = m$, $\forall i$ (see
figure~\ref{fig:3equalmasses} for the $N=3$ case).  Then the critical
points of the potential eq.(\ref{eq:V(phi)}) satisfy $\phi_i = \phi$ or $\pi-\phi$ $\forall i$, for some angle $\phi$ which is determined by the unitarity constraint. 

At a local minimum, all $\phi_i$ are equal 
\beq\label{eq:equalphi}
\phi_i  = \frac{2\pi n - \thb}{N} \equiv \phi \qquad \forall i~,
\eeq
and have postive cosine, so that $n \in {\mathbb Z}$ satisfiesy
\beq\label{eq:ncond}
-\frac{N}{4} + \frac{\thb}{2\pi} < n < \frac{N}{4} + \frac{\thb}{2\pi} ~.
\eeq   
Note that if the boundary values satisfy $\thb/2\pi\pm  N/4 \in {\mathbb Z}$ this does not reliably lead to extra metastable states.
This is because the eigenvalues of the Hessian at a local min are $N m \cos\phi$ (once) and $m \cos\phi$ (with multiplicity $N-2$), which all vanish at such a point and so a higher-order analysis including the $\OO(M_q^2)$
and electromagnetic terms in the chiral Lagrangian is necessary to determine if there are shallow local minima in this marginal case.  
So, to be conservative, we only count those minima strictly satisfying the condition eq.(\ref{eq:ncond}).

In any case, the value of the potential at the local minima specified by eq.(\ref{eq:equalphi}) subject to eq.(\ref{eq:ncond}) are
\beq\label{eq:equalpot}
V^{(n)} = - N m B_0 \cos\left(\frac{2\pi n - \thb}{N}\right)~,
\eeq
so if metastable local minima exist they are typically split in energy density from the ground state by $\OO(m\Lambda_{qcd}^3)$, apart from the exceptional
case of $\thb=\pi$ where there are two degenerate states \cite{Smilga:1998dh}.  Except in tuned cases, $\OO(m\Lambda_{qcd}^3)$ is also a good estimate of the splitting between the metastable states and the ground state in the general case of unequal masses if the mass parameter is chosen to be that of the lightest non-zero quark mass.  

It is amusing to count the number of local minima in this equal quark mass case as $N\geq 2$ and $\thb$ vary.  For example one finds that at $\thb=0$ the total
number of reliably predicted metastable minima, $N_s$ (excluding the true ground state), is given as
\beq\label{eq:numbervacua-equal}
N_s (\thb=0) = 2[N/4]_<
\eeq
where $[x]_<$ denotes the integer part of $x$ with magnitude strictly less than $|x|$.  So for $N=2,3,\ldots,6$ one finds $N_s=(0,0,0,2,2)$, and thus that
five equal mass light quarks is the first case one definitely finds metastable vacua at $\thb=0$.   On the other hand
for, eg, $\thb=\pi/2$,  $N_s = [(N+1)/4]_< +[(N-1)/4]_<$, giving $N_s=(0,0,1,1,2)$, so metastable vacua definitely first appear when there are
four equal mass light quarks.

The tunnelling rate between these metastable states and the ground state (or a lower metastable state if there are many) has been calculated in simple limits in Refs.~\cite{Smilga:1998dh,Dubovsky:2010je}.  For example, just away from $\thb=\pi$ there are two slightly split vacua with a difference in energy densities given by
$\Delta V \sim \sqrt{3}  B_0 m |\delta|$ where $\thb=\pi+\delta$ and $|\delta| \ll 1$ is assumed.  Then the result of a thin-wall false vacuum decay calculation gives
a zero-temperature decay rate per unit volume \cite{Smilga:1998dh} (neglecting sub-leading $\OO(1/|\delta|^2)$ terms in the exponent)
\beq\label{eq:decayrate}
\frac{\Gamma}{\rm Vol} \sim \Lambda_{qcd}^4 \,  \exp\left( -D \frac{f_\pi^4}{ m B_0 |\delta|^3}\right)~,
\eeq
with a numerical factor $D\simeq 4\times 10^3$.  Because of this large numerical factor, even for
$\delta \not\ll 1$ (as long as $\delta$ does not approach $\pi/2$ where the metastable state ends) and relatively `heavy' light quarks with $m \sim \Lambda_{qcd}$,
the metastable state is predicted to be long lived in this equal quark mass case.

\subsection{Fan diagrams}

Away from the special case of equal masses the analysis is more complicated and analytic formulae are not particularly illuminating.
Fortunately, as already indicated, the general conditions in section \ref{sec:criticalpoints} lend themselves to a diagrammatic interpretation of
critical points, a `fan' diagram, which is particularly useful for the qualitative identification of metastable states for any $\{m_i,\thb\}$.

\begin{figure}[t]%
\centering
\subfigure[\emph{Critical point}]{\label{fig:critical_points}\includegraphics[width=0.495\linewidth]{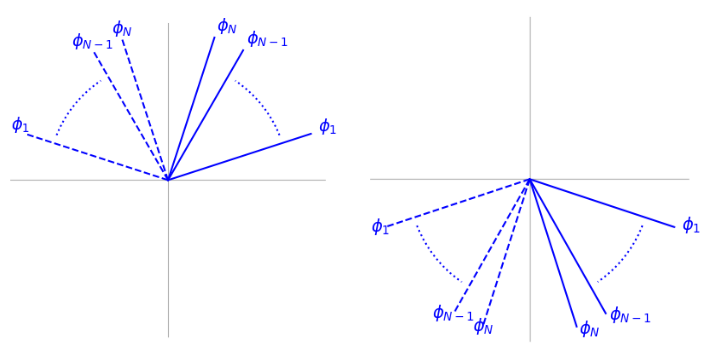}}
\subfigure[\emph{Local minima}]{\label{fig:local_minima}\includegraphics[width=0.495\linewidth]{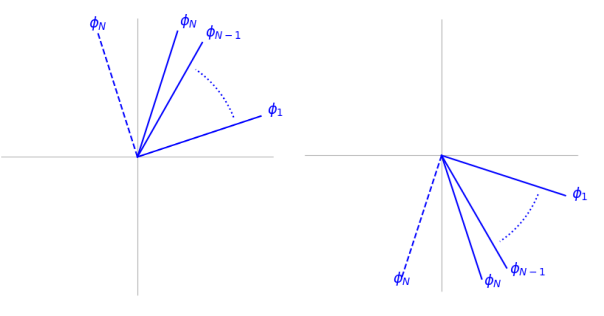}}%
\caption{Necessary qualitative arrangements of the $\phi_i$ at a general critical point and a local minimum. The `spreadout-ness' of these `fan diagrams' is determined by ratios $m_i/m_j$
via eq.(\ref{eq:sin_idts}). As $m_N$ was chosen to denote the lightest
quark,  the angle $\phi_N$ will always be closest to the $y$ axis, with successive angles flattening out towards the $x$ axis with separations determined by the ratios $m_i/m_j$ (if $m_i = m_j$ then $\phi_i =\phi_j \; \text{or} \;\pi - \phi_j$). Dashed lines in figure~\subref{fig:critical_points} denote an equally possible alternative for each individual angle. Thus there are $2 \cdot2^N$ qualitativelty different arrangements that may constitute a critical point. Generally, only a subset of these will be consistent with the constraint $\sum_i \phi_i = -\thb $ mod $2\pi$, which determines the absolute value of the $\phi_i$.
 The extra requirement of a local minimum picks out the arrangements in figure~\subref{fig:local_minima}. Here, for the dashed $\phi_N$ to be a valid alternative for a local minimum the extra constraint eq.(\ref{eq:tan_constraints}) must also be satisfied.}
\end{figure}

A local minimum, if it exists, corresponds to an ordered arrangement of angles as in figure~\ref{fig:local_minima}, where the spread between angles is fixed by the respective mass ratios according to eq.(\ref{eq:sin_idts}), so as the mass ratios approach unity the separation angles between the corresponding elements of the fan become smaller, and conversely increase as the mass ratios increase.

\begin{figure}
\centering
  \includegraphics[scale=0.6]{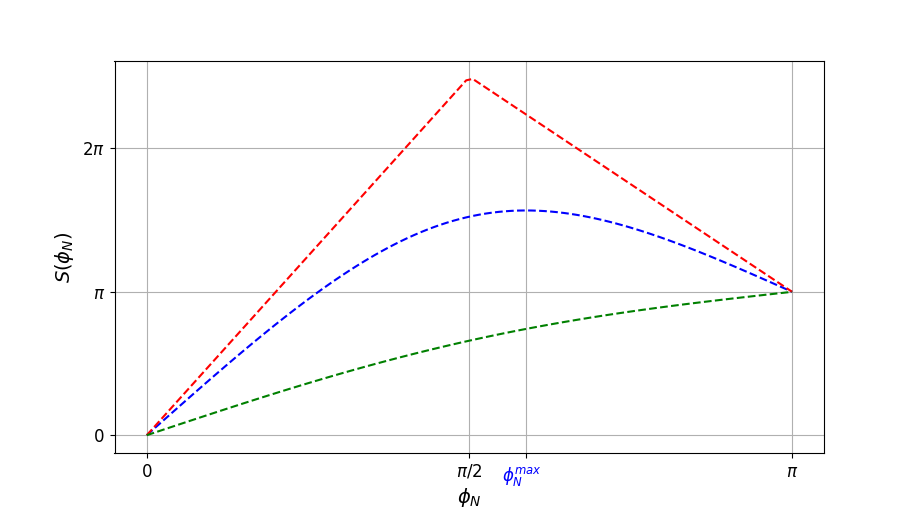}
\caption{The sum of angles $S$ as a function of leading angle $\phi_N \in (0,\pi)$ for different values of quark masses. $S(\phi_N)$ on $\phi_N \in (-\pi,0)$ is inferred, being an odd function. If $S$ hits the target angle $-\thb$ mod $2\pi$ while its slope is positive then we have a local minimum. The position of the maximum value of $S(\phi_N)$ is mostly controlled by the two lightest quark masses. The red curve is characteristic of  $m_N = m_{N-1}$, when $\phi_N^{max}=\pi/2$ always. As their ratio increases $S(\phi_N)$ goes from having $\phi_N^{max}\in(\pi/2,\pi)$ (blue line) to $\phi_N^{max}=\pi$, achieved at the boundary (green line). \label{fig:S_of_thetaN}} 
\end{figure}

To find a local minimum one adjusts the value of the `leading' angle $\phi_N$ (recall that $m_N$ is the lightest quark), from which all other (subordinate) angles follow via $\phi_i=\sin^{-1}\left(\frac{m_N}{m_i}\sin\phi_N\right)$, until the sum of all $\phi_i$ hits the `target' angle demanded by the unitary condition 
\beq
\label{eq:sum_of_angles}
S\equiv \sum_{i=1}^N \phi_i = -\thb \quad {\rm mod} \;2\pi~.
\eeq
If the target angle is hit for $|\phi_N|<\pi/2$ the configuration is guaranteed to be a local minimum. Otherwise ($\cos \phi_N<0$) one also requires eq.(\ref{eq:pos_def_function}) to ensure minimality, which in this case can be reformulated in a version suited to diagrammatic analysis that only depends on the angles 
\begin{equation}
\label{eq:tan_constraints}
|\tan(\phi_N)| > |\tan(\phi_1)| + \dots + |\tan(\phi_{N-1})|, \quad \left(\frac{\pi}{2}< \phi_N < \pi\right)~.
\end{equation}
Importantly, fan diagrams are useful because \emph{if for a given set of parameters}, $\{y_\alpha\}$, \emph{two distinct local minimum fan diagrams can be drawn, the theory admits metastable states.}

For a quantitative handle it is useful to consider the sum in eq.(\ref{eq:sum_rate}) as an explicit function of the leading angle $\phi_N$
\begin{equation}
\label{eq:angles_sum}
	S(\phi_N) = \sin^{-1}\left(\frac{m_N}{m_1}\sin\phi_N\right) + \dots + \sin^{-1}\left(\frac{m_N}{m_{N-1}\sin\phi_N}\right) + \phi_N~.
\end{equation}
For all quark masses, $S$ goes from the `closed fan' configuration $S(0)=0$ to $S(\pi)=\pi$. Clearly $S(\phi_N)$ is monotonically increasing for $\phi_N \in (0,\pi/2)$ as all other subordinate angles also become larger. 
After $\phi_N>\pi/2$, however, the latter decrease as $\phi_N$ increases: the behaviour of $S(\phi_N)$ depends on the mass ratios at hand.
Insight is obtained by differentiating eq.(\ref{eq:angles_sum}),
\begin{equation}
\label{eq:sum_rate}
	S'(\phi_N) = C(\phi_N)~,
\end{equation}
where $C(\phi_N)$ was the function whose positivity implied a local minimum, eq.(\ref{eq:pos_def_function}), equivalent to eq.(\ref{eq:tan_constraints}). Thus, the mark of a local minimum is equivalently expressed as
\begin{equation}
	S(\phi_N) = -\thb \quad {\rm mod} \;2\pi, \quad S'(\phi_N) >0.
\end{equation}

Note also that $C'(\phi_N)<0$ on $(0,\pi)$. Depending on the values of $\{m_i\}$, $S(\phi_N)$ will either always monotonically increase or achieve a maximum $\in (\pi/2, \pi)$ as shown in 
figure~\ref{fig:S_of_thetaN}. 

Notice that here our primary concern is the \emph{existence} of (multiple) local minima, rather than their specific location. For any $\theta \in (-\pi,\pi)$ there will always be a first local minimum with $|S(\phi_N^{max})|=|\theta|$. Suppose $S$ achieves its largest value at $\phi_N^{max}$. A sufficient condition for the existence of a second minimum is
\begin{equation}
\label{eq:existence_condition}
	S(\phi_N^{max}) > 2\pi-|\thb|~,
\end{equation}
which ensures the target angle is hit a second time \emph{for some value of $\phi_N$}.

\subsection{Metastable states at $\thb = 0$}\label{sec:theta0}

For $\thb = 0$, the global minimum of the potential corresponds to the trivial `closed fan' arrangement $\phi_i = 0$ $\forall i$. We will have metastable states if there
exist other arrangements satisfying
\begin{equation}
	S(\thb_N) =  2\pi n,  \quad \quad S'(\phi_N)>0~,
\end{equation}
with $ n\in {\mathbb Z}/\{0\}$.
 These will always come \emph{in pairs} as for $\thb =0$ we have a $\phi_j \rightarrow - \phi_j$ symmetry. 
 We focus on diagrams with positive angles, unless otherwise stated. 

It is natural to start from the case of all masses equal $m_i = m_j$, implying $\phi_i = \phi_j$ for all $i,j$, already discussed in section \ref{sec:equalmasses}. Even with the maximum values $\phi_i=\pi/2$ clearly no appropriate fan diagram may be drawn for $N\leq 3$. For $N=4$, the diagram $\phi_{1-4} = \pi/2$ saturates eq.(\ref{eq:pos_def}) and so is sensitive to higher order
corrections to our effective Lagrangian eq.(\ref{chiralL}), and a deeper analysis is required.  

\begin{figure}[t]
\centering
\begin{floatrow}
  \ffigbox[\FBwidth]{ \caption{QCD at $\thb =0$ with 5 equal mass light quarks has a metastable state (blue) as well as the global minimum (red). There is also another (degenerate in $V$) metastable state corresponding to $\phi_{1-5} =-2\pi/5$. This behaviour persists for mass ratios satisfying $S(\phi^{max}_5\{m_i\})>2\pi$.}  \label{fig:N5_sol} }{
    \includegraphics[width=\linewidth]{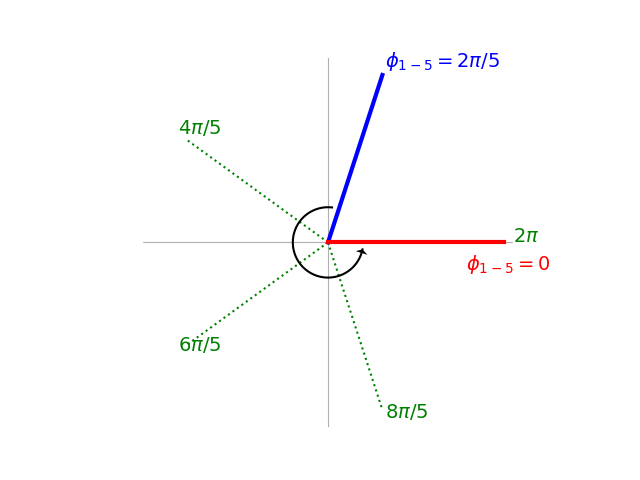}   
  }
  \ffigbox[\FBwidth]{\caption{Global minimum (red) and metastable state (blue) in a 6 light quark theory with $\thb = 2\pi/5$ and mass ratios $10:2:1.5:1.3:1.1:1$. In general, for high $N$ and certain mass ratios, there may also be more minima, corresponding to diagrams whose angle sum makes an extra complete revolution before hititng the target. } \label{fig:6lqThetaFan} }{%
    \includegraphics[width=\linewidth]{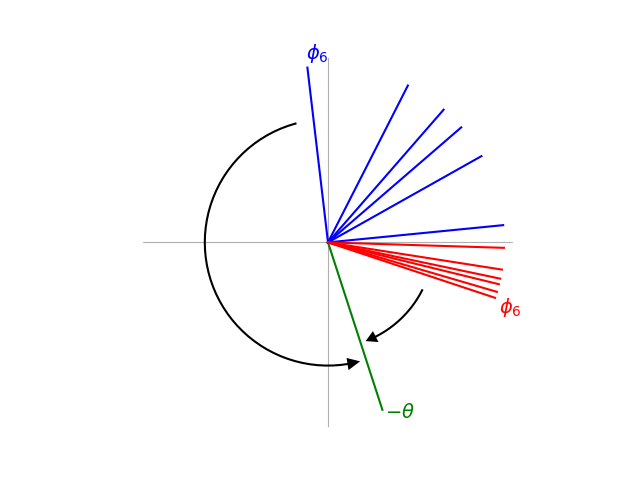}  
  }
\end{floatrow}
\end{figure}

As already mentioned, $N=5$ is the first case with metastable states for $\thb =0$. For all masses equal, this corresponds to the diagram in figure~\ref{fig:N5_sol}.  Analogous diagrams can be drawn for $N>5$, explicitly with $\phi_i = 2\pi/N$. Moving away from the equal mass theory point, there will metastable states for a range of non-trivial mass ratios, corresponding to spreading of the angles in figure~\ref{fig:N5_sol}. For example, if the lightest four quarks are degenerate, then it should be clear from the diagram that the remaining one can be arbitrarily heavier (within the range of validity of the chiral Lagrangian): $\phi_1$ can be arbitrarily small, compensated by $\phi_{2,3,4,5}$ moving arbitrarily close to $\pi/2$. A full chart of the swamp (by measure of the SdSC) for arbitrary $\{N,m_i\}$ is identified with the region satisying\footnote{See eq.(\ref{eq:existence_condition}) and surrounding text.} 
\begin{equation}
 	S(\phi_N^{max}(m_i))>2\pi  \quad \quad (\thb =0)~,
\end{equation} 
where the dependence of $\phi_N^{max}$ on $\{m_i,N\}$ is non-trivial, though easily implementable numerically. A good approximation is obtained by replacing $\phi_N^{max} \rightarrow \pi/2$ as suggested in figure \ref{fig:S_of_thetaN}, which gives
\begin{equation}
\label{eq:guarantee_theta_zero}
	\sin^{-1}\left( \frac{m_N}{m_1} \right) + \dots + \sin^{-1}\left( \frac{m_N}{m_{N-1}}\right) \gtrsim \frac{3\pi}{2} \quad \quad (\thb =0)~.
\end{equation}
The latter becomes an exact condition in the limit $m_{N-1} \rightarrow m_N$, is an underestimate when $m_{N} \lesssim m_{N-1}$ and again valid when $m_{N} \ll m_{N-1}$ (in which case there are no metastable states).

\subsection{Non zero $\thb$-angle}

As $\thb \in (-\pi,\pi)$ turns on, the global minimum of the potential moves away from the closed fan to a fan diagram
with $\text{sign}(\phi_i)=-\text{sign}(\thb)$.  At least one metastable state will be present if there exist also an arrangement with $\text{sign}(\phi_i)=\text{sign}(\thb)$
as in the example in figure~\ref{fig:6lqThetaFan}. Clearly, the bigger $\thb$ is, the larger the space of mass ratios with metastable states is, as
less `angular power' is necessary to reach the target angle $-\thb$. This continues until $\thb =\pi$, when global and first metastable branches become degenerate and in the neighbourhood of which
the space of masses $\{m_i\}$ admitting a metastable state is at its largest, set by\footnote{This can be obtained by noting from figure \ref{fig:S_of_thetaN} that there can be no local minimum other than the global one in the neighbourhood of $\thb=\pi$ iff  $S'(\pi) = C(\pi)>0$.}
\begin{equation}
\label{eq:safe_zone}
	\frac{1}{m_N} < \frac{1}{m_1} + \dots + \frac{1}{m_{N-1}} \quad \quad (\thb=\pi)~.
\end{equation}

\begin{figure}[t]
\centering
\begin{floatrow}
  \ffigbox[\FBwidth]{ \caption{For three light quarks, regions with metastable states in the mass ratio plane for different values of $\thb$ according to the leading order
 chiral Lagrangian analysis.  Our world, with the values of $m_u=2.16^{+0.49}_{-0.26} \MeV$, $m_d = 4.67 ^{0.48}_{-0.17} \MeV$ and $m_s=93^{+11}_{-5} \MeV$ \cite{Review_of_PP} is indicated by the red `dot'. \label{fig:3lqTheta_arrow}} }{
    \includegraphics[width=\linewidth]{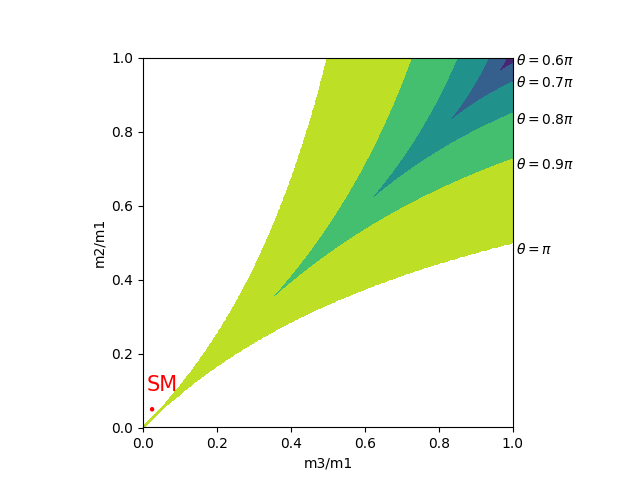}   
  }
  \ffigbox[\FBwidth]{\caption{For four light quarks, fixing $m_{1}=m_{2}$ and $m_{3}=m_{4}$, the region in red, bordered by the equation in figure, corresponds to theory points with metastable states according to the leading order chiral Lagrangian analysis. } \label{fig:4lqTheta} }{%
    \includegraphics[width=\linewidth]{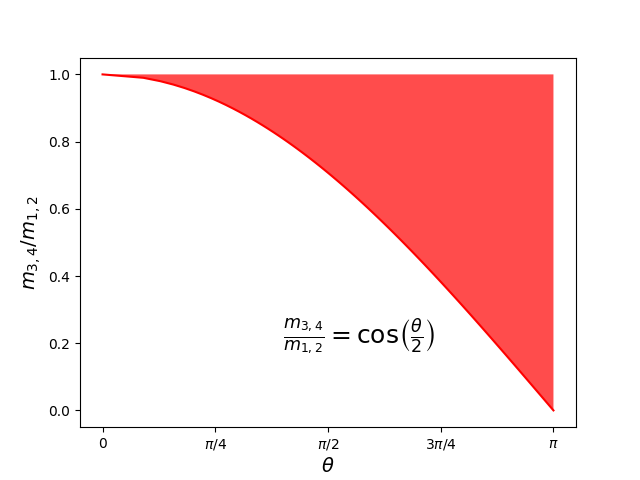}  
  }
\end{floatrow}
\end{figure}

In particular, the case of \emph{two light quarks} remains safe from metastable states for any mass ratio $m_1/m_2$ and vacuum angle $\thb$.
The case of \emph{three light quarks} (our world) starts showing metastable states after $\thb > \pi/2$, starting from the equal masses theory point and expands from there as shown in
figure~\ref{fig:3lqTheta_arrow}. It is interesting to note that the SM ratio of light quark masses satisfies
\begin{equation}
	\frac{1}{m_u} > \frac{1}{m_d} + \frac{1}{m_s},
\end{equation}
where $m_{u,d,s}$ are the up, down and strange quark masses respectively,
making the theory `safe' from metastable states \emph{for all values of} $\thb$.
As more quarks are made light, metastable states are encountered more and more frequently.  Figure~\ref{fig:4lqTheta} shows a particular section of parameter space for four light quarks, while figure~\ref{fig:Vbranches} shows a similar slice for the case of six light quarks. Once again, we can write a condition mapping out the swamp (by measure of the SdSC) as
\begin{equation}
	S(\phi_N^{max}(m_i)) > 2\pi-|\thb|~,
\end{equation}
which, in its approximated $\phi_N^{max}\rightarrow \pi/2$ form, appeared as eq.(\ref{eq:master_eq}).

Finally, it is also noteworthy that the theory points at the centre of the regions of metastable states correspond to the lightest quarks being degenerate (equal masses), the ensuing global symmetry thus simultaneously violating another swampland conjecture.

\section{\label{quint}Coupling to quintessence}

The swampland de Sitter conjecture has immediate relevance to our present environment; we must not currently live in a positive energy density local minimum.  As a
result, the cosmological constant is eponymously defiant, in that it must not be constant!  As we briefly discussed in the Introduction this suggests (but does not uniquely imply) that there is currently some quintessence scalar $\varphi(t)$ contributing a positive energy density which is slowly evolving with time.  

It is striking that, when confronted with observed cosmological parameters, simple models of quintessence are on the cusp of tension with the SdSC \cite{Agrawal:2018own}.  In particular, eqs.(\ref{eq:rdSc1}) and (\ref{eq:rdSc2}) imply that either the slope of the potential in the $\varphi$-direction around its present value, $\varphi_0$, is bounded below by an amount that depends on the value of the total effective vacuum energy now, $V(\varphi_0,\ldots)\equiv V_0>0$, or $\varphi_0$ is a local maximum in the $\varphi$-direction with curvature around that point being similarly bounded.  But the quintessence field must not be so fast evolving that within a fraction of a Hubble time a deep anti-de-Sitter state is reached, and an associated ``big crunch'' occurs.\footnote{Also in our particular Universe one must be consistent with the observational bound on the deviation of the equation of state parameter from $\omega \simeq -1$, though we will \emph{not} need to use this tighter constraint.}  

Since quintessence may render the known Universe consistent with the swampland conjectures it is natural to consider whether it may also pose a potential
significant caveat to the inconsistency of metastable QCD states with the conjectures.  To this end, let us consider two possibilities
\begin{enumerate}[(i)]
\item Sequestered quintessence:  $V\approx V_{\{ y_\alpha\},\text{QCD}}^{(n)}(\pi^a)+ {\tilde V}(\varphi)$
\item Coupled quintessence:  $V= V_{\{ y_\alpha\},\text{QCD}}^{(n)}(\pi^a,\varphi)$
\end{enumerate}
where as before $n=0,1,\ldots$ labels all the branches.  In the second case there is a non-trivial
coupling between the QCD and quintessence sectors, and thus also between $\varphi$ and the SM more generally.  For either case there are two further
considerations.  The first is whether or not a scenario is consistent with the swampland conjectures.  The second is whether or not it may be 
consistent with having a Universe that avoids a big crunch occurring on an extremely short timescale.

\subsection{Sequestered quintessence}

\begin{figure}
\centering
  \includegraphics[scale=0.3]{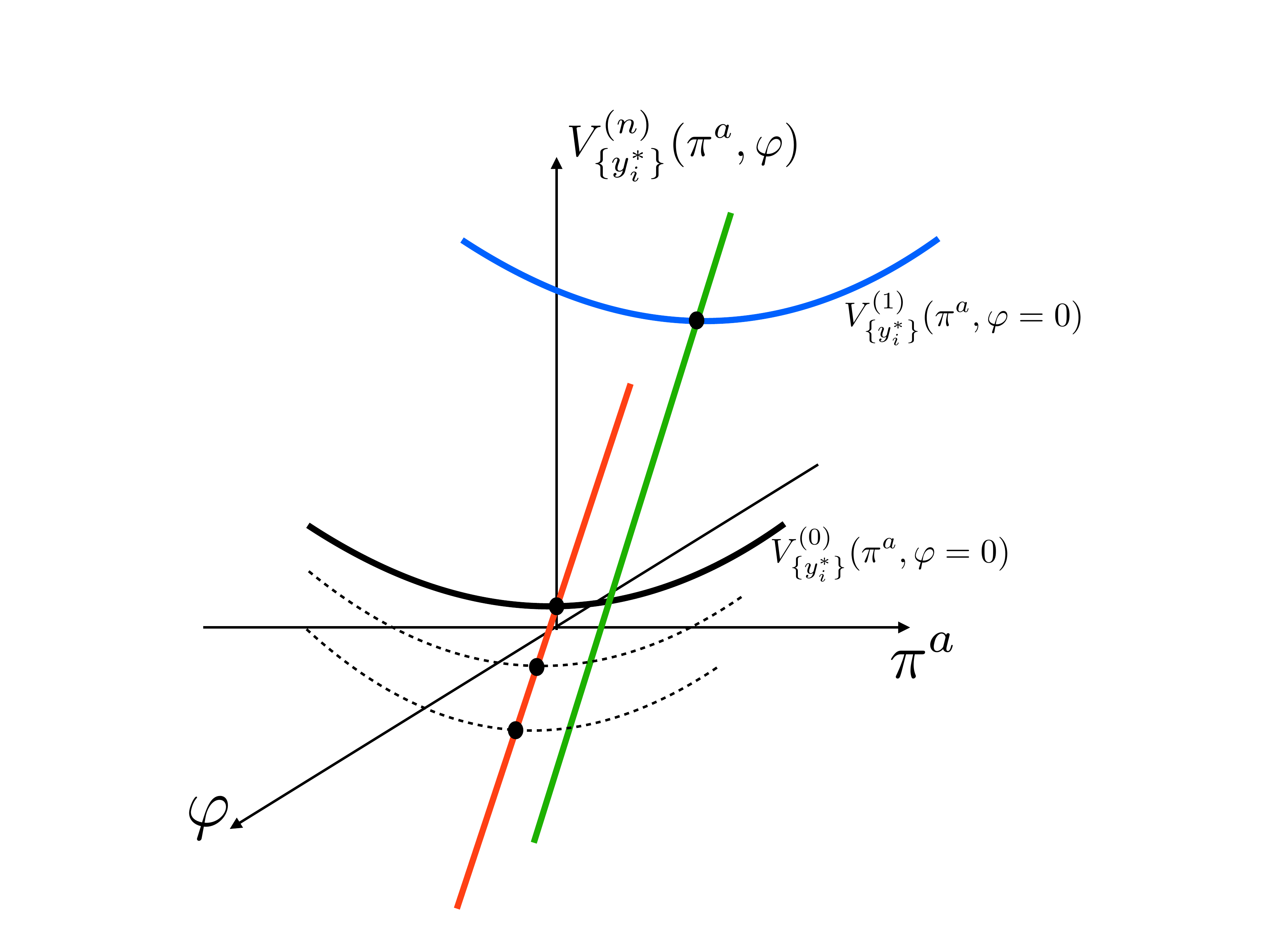}
\caption{Schematic illustration of sequestered quintessence.  The black curve gives the form of
the ground state branch as a function of the vev's of pNGB fields $\pi^a$ at a value of the quintessence field
$\varphi=0$, while blue curve shows form of first metastable branch possessing minimum (in absence of $\varphi$) shifted in $\pi^a$.  Both curves
at are same values of all SM parameters $\{ y_\alpha^* \}$. The green curve shows bottom of the valley in the $\varphi\neq 0$ direction
starting from the otherwise-metastable minimum, while the red curve shows same for valley starting from true ground state.  For sequestered
quintessence the slope of the red and green trajectories is the same.  The SdSC demands that the slope (or curvature) of the green
curve is bounded below by an amount proportional to the effective positive vacuum energy density of the would-be metastable state. \label{fig:quint1}} 
\end{figure}

\begin{figure}
\centering
  \includegraphics[scale=0.3]{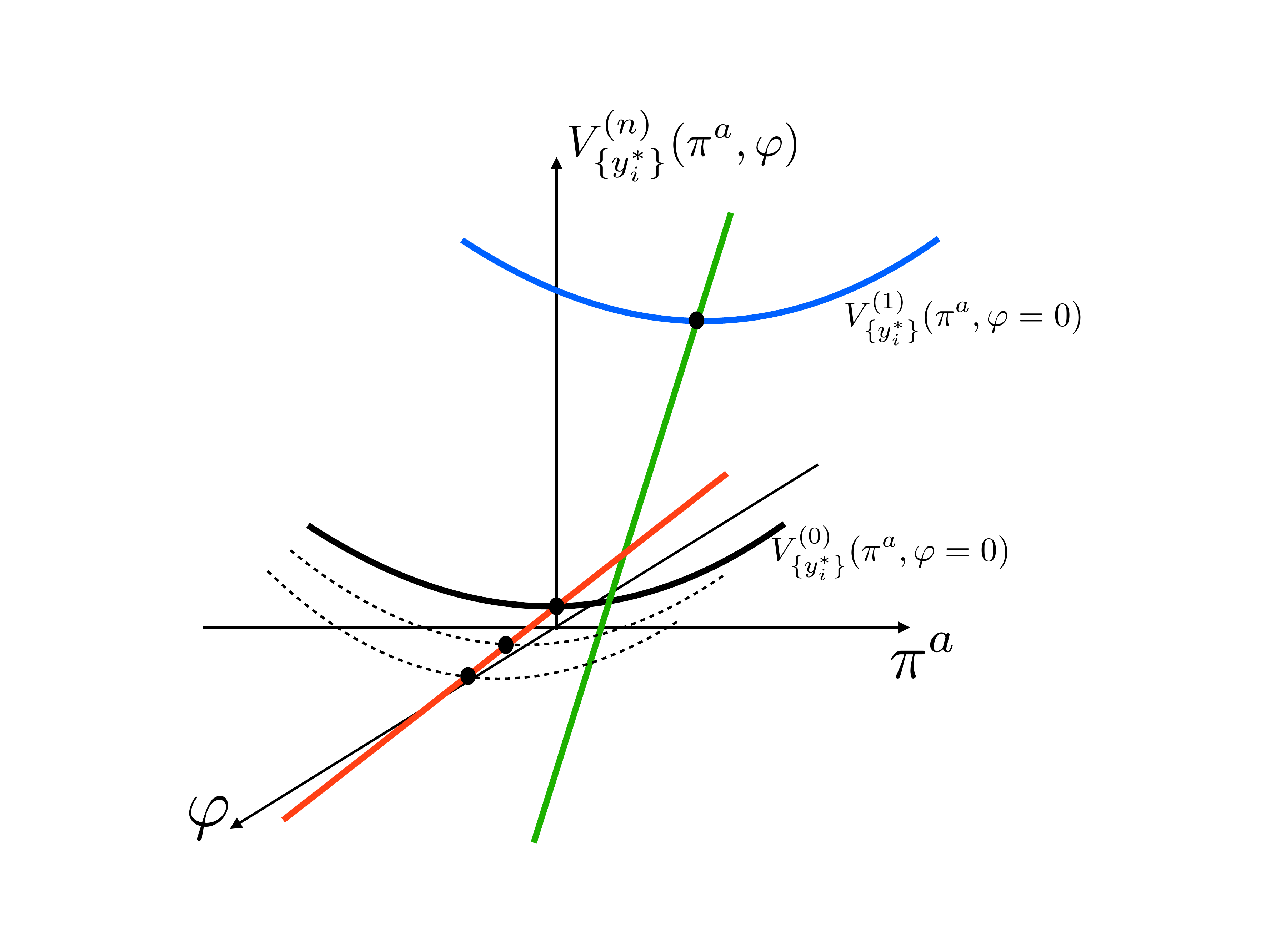}
\caption{Same as figure~\ref{fig:quint1} except for coupled quintessence.  In this case the slope and curvature of the quintessence valley
sloping away from the ground state of QCD differs from that of the metastable valley.  In principle this could allow the green curve valley
bottom to satisfy the SdSC while the red valley bottom is not so steep as to lead to immediate evolution to an AdS state.  However such a large
change in slope between the two valleys implies a large coupling between the SM and $\varphi$ giving an extreme fine-tuning
in the quintessence sector.\label{fig:quint2}} 
\end{figure}

The case of sequestered quintessence is straightforward, as the slope and curvature of the potential in the $\varphi$ direction are independent
of whether the QCD sector is in the ground state or a metastable state.  This situation is illustrated in figure~\ref{fig:quint1}, while a case of
coupled quintessence is sketched in figure~\ref{fig:quint2}.  Applying the SdSC conditions to the metastable state at SM parameter
values $\{ y_\alpha^*\}$ with large and positive vacuum energy $\Delta V_{\{ y_\alpha^*\},\text{QCD}} \sim m_q \Lambda_{qcd}^3$ (here $m_q$ is an appropriate
quark mass) gives
\beq\label{eq:quint_cond}
|\nabla_\varphi  {\tilde V}(\varphi)|  \geq   c \frac{m_q \Lambda_{qcd}^3}{\mpl} ~~~~
{\rm or}~~~~\nabla_\varphi^2 {\tilde V}(\varphi)   \leq   -c' \frac{m_q \Lambda_{qcd}^3}{\mpl^2} 
\eeq
where on the RHS of both inequalities we have dropped the tiny corrections from $V_0 \ll m_q \Lambda_{qcd}^3$.  But the sequestered form
now implies that in the QCD ground state branch at $\{ y_\alpha^*\}$ the slope or curvature is just as large.  

Specifically, in the case that the first, slope, condition of eq.(\ref{eq:quint_cond})
is satisfied, the $\varphi$ equation of motion for the ground state branch is (dropping the Hubble friction term, $3H \dot\varphi$, as the
vacuum energy, $V_0$, and thus $H$ is now tiny)
\beq
\ddot \varphi(t) \gsim c \frac{m_q \Lambda_{qcd}^3}{\mpl} 
\eeq
implying that in a time $\Delta t \equiv \tau/H$ (here we measure time in fractions of the Hubble time $H^{-1} \sim \mpl/\sqrt{V_0}$) the quintessence field
evolves by an amount $\Delta \varphi \sim \tau^2 c \mpl m_q \Lambda_{qcd}^3 / V_0\gg \mpl$ unless $\tau\ll 1$ since $m_q \Lambda_{qcd}^3\gg  V_0$. 
Here we have made the conservative assumption that the initial quintessence field velocity, $\dot\varphi=0$. (Note
that the Swampland Distance Conjecture \cite{Ooguri_2007,Ooguri:2018wrx} states that the effective field theory describing the quintessence plus SM
system must irrevocably break down once $\Delta\varphi \gsim \mpl$.  This limits $\tau \lsim  (V_0/m_q\Lambda_{qcd}^3)^{1/2}$.) During this evolution
the value of the vacuum energy density in the ground state branch becomes
\beq
V(\varphi_\tau,\ldots) \simeq  V_0 - c^2 \tau^2 \frac{m_q^2 \Lambda_{qcd}^6}{V_0}~.
\eeq
Thus as $c$ is $\OO(1)$, within a fraction 
\beq
\tau_{AdS}\sim  \frac{V_0}{m_q \Lambda_{qcd}^3} \ll 1
\eeq
of a Hubble time the Universe evolves to an anti-de-Sitter state.   Taking as an example, the present inferred value of the
vacuum energy $V_0 \simeq 10^{-47}\GeV^4$, and, conservatively,
$m_q = m_u\simeq 3\MeV$ gives $\tau_{AdS}\simeq 10^{-43}$, so we would collapse essentially instantaneously!  Only for truly exponentially small
$m_q$ and thus exponentially small explicit breaking of the chiral symmetry can we possibly avoid this catastrophe.\footnote{It is amusing to contemplate
that there is an interplay between this exponentially small value of explicit chiral symmetry breaking, the quantitative lower bounds on explicit
global symmetry violation imposed by the Swampland Global Symmetry Conjecture \cite{Fichet:2019ugl,Daus:2020vtf}, and the size of the vacuum energy $V_0$.}

Alternatively the quintessence field $\varphi$ might be located very close to a local maximum of the potential, in which case this is a
form of ``hilltop quintessence" \cite{Dutta_2008}.  Suppose that the initial displacement of the field $\varphi$ from the exact maximum is $\delta\varphi$, and again
conservatively assume that the initial field velocity $\dot\varphi=0$.
One then finds using the same logic as in the previous paragraph that the curvature SdSC condition applied to the metastable branch implies for the ground
state branch that the dimensionless timescale (fraction of a Hubble time) until evolution to an AdS state is now set by
\beq
\tau_{AdS}\sim  \frac{V_0}{m_q \Lambda_{qcd}^3} \log\left(\frac{V_0 \mpl^2}{(\delta\varphi)^2 m_q \Lambda_{qcd}^3}\right) .
\eeq
This is also an extremely short timescale unless the initial displacement $\delta\varphi$ from the maximum of the potential is tuned
to be super-exponentially small.  (Moreover, in a cosmological context, there are automatically fluctuations in $\varphi$
entering the horizon, displacing the value of the quintessence field from the maximum of the potential.) 

Thus no matter which of the SdSC conditions applies, the presence of a sequestered quintessence sector does not allow one to evade the SdSC
constraint on SM parameter values $\{ y_\alpha^*\}$ coming from the presence of metastable QCD-sector states.  The fundamental reason for this is of course that the 
the difference in effective vacuum energy between the ground state and the metastable state(s) at a particular value of the SM parameters $\{ y_\alpha^*\}$ is 
large, and it is impossible to satisfy the constraint on the potential in the quintessence field direction for the metastable state without leading
to catastrophic evolution of the ground state.  

\subsection{Coupled quintessence}

One can see that this is not just a feature of the sequestered quintessence potential: To simultaneously 
satisfy the SdSC constraint in the metastable state and avoid the far-too-steep or curved potential in the ground state requires that the form
of the potential in the $\varphi$-direction very significantly changes as one moves from the metastable branch at $\{ y_\alpha^*\}$ to the ground state
branch at the same $\{ y_\alpha^*\}$ (see figure~\ref{fig:quint2}).  But the only difference between these two branches is the presence of pNGB vacuum expectation values of
typical size $\sim f_\pi$.  For the quintessence potential function to change very greatly in slope or curvature in response to the switching on of
these pion vev's then requires that the quintessence field couples \emph{significantly} to QCD.  Parametrically one needs 
\beq
\partial_{\pi^a}^2\partial_\varphi  V_{\{ y_\alpha\},\text{QCD}}^{(n)}(\pi^a,\varphi) \sim \frac{m_q \Lambda_{qcd}^3}{\mpl f_\pi^2}~,
\eeq
which then implies that the effective cubic $\epsilon\varphi (\pi^a)^2$ coupling is $\epsilon \sim m_q \Lambda_{qcd}^3/\mpl f_\pi^2 \sim m_q \Lambda_{qcd}/\mpl$.  Similarly
the quartic coupling $\lambda \varphi^2 (\pi^a)^2$ is given by $\lambda \sim m_q\Lambda_{qcd}^3/\mpl^2 f_\pi^2 \sim m_q \Lambda_{qcd}/\mpl^2$.  There are also of course
couplings to higher powers of the pNGB fields as implied by the non-linearly realised approximate chiral symmetry.  These couplings between the SM and 
$\varphi$ imply that even in the ground state branch of QCD there are, upon integrating out the pions (with mass $m_pi^2 \sim m_q f_\pi$), radiative corrections to, eg, the mass-squared of $\varphi$ of size
$\delta m^2_\varphi \sim (m_q \Lambda_{qcd})^2/16\pi^2 \mpl^2$.  These are enormous compared to the mass $m_\varphi \sim H_0 \sim \sqrt{V_0}/\mpl$ required of a
successful quintessence field.  Similarly other parameters of the quintessence potential get very large correction.  In other words, one must exponentially tune by an amount $\sim (m_q \Lambda_{qcd})^2/16\pi^2 V_0$ (in our Universe roughly a 1 part in $10^{39}$ tuning) $m_\varphi^2$, as well as other parameters,  to get the coupled quintessence model not to again evolve very quickly to AdS and a big crunch.

Thus the presence of a quintessence field does not obviously invalidate the arguments concerning the inconsistency of metastable QCD states with the SdSC. 
We emphasise that in these arguments we have not at all used the observational or experimental constraints on quintessence fields, such the equation of state parameter
bounds, or the limits on fifth-forces or equivalence-principle violation, which of course only apply to our Universe.  We have only used the fact that the SdSC must, by assumption, be satisfied for the metastable states, together with the necessity of having the putative Universe in a QCD ground state (at the same values of the SM parameters) survive longer than an exponentially short period of time before entering an AdS big crunch phase.

\section{Standard Model with $N=0$ light quarks and the $v_{EW}\rightarrow \infty$ limit}\label{sec:N=0}

A question prompted by the proceeding analysis is if the $N=0$ light quark limit is similarly constrained by the SdSC.  There are two ways one can
achieve this limit.  Either all the quark Yukawa couplings can be taken to be $\gg 0.01$ so there are no quarks with masses less than $\sim 4\pi f_\pi\simeq 1 \GeV$.
Or the Yukawa couplings can remain at their standard values while the electroweak vacuum expectation value (vev) is taken to be $\gg 50 \TeV$.  In the second
case our thinking then possibly leads to a new perspective on the hierarchy problem.  

Let us first consider the extreme limit of the SM where the EW vev,  $v \rightarrow \infty$, with all Yukawa and gauge couplings kept fixed as well as the higgs quartic coupling.  In this limit the IR theory is of course a pure $SU(3)\times U(1)_{EM}$ gauge theory with no matter, as all quarks, leptons, weak gauge bosons, and the higgs boson itself
have become super massive.   The question is if this theory has metastable states.  

Because there is no matter, the color and EM gauge groups are completely decoupled from each other, so the question becomes do either, or both, of pure $U(1)$ and $SU(3)$ theories have metastable states.  There is no argument that we are aware of that indicates that $U(1)$ has metastable states, but the situation is plausibly different for $SU(3)$.

\subsection{Metastable states of pure $SU(N_c)$ gauge theories}

Following early suggestions \cite{Witten:1980sp}, Witten \cite{Witten:1998uka} and Shifman \cite{Shifman:1998if} argued that for large $N_c$, non-supersymmetric pure $SU(N_c)$ possesses $N_c-1$ metastable vacua, as reviewed in Ref.\cite{Teper:2008yi}.\footnote{We particularly thank Mike Teper for discussions of the status of metastable states in $SU(N_c)$ theories.} 

\begin{figure}
\centering
  \includegraphics[scale=0.3]{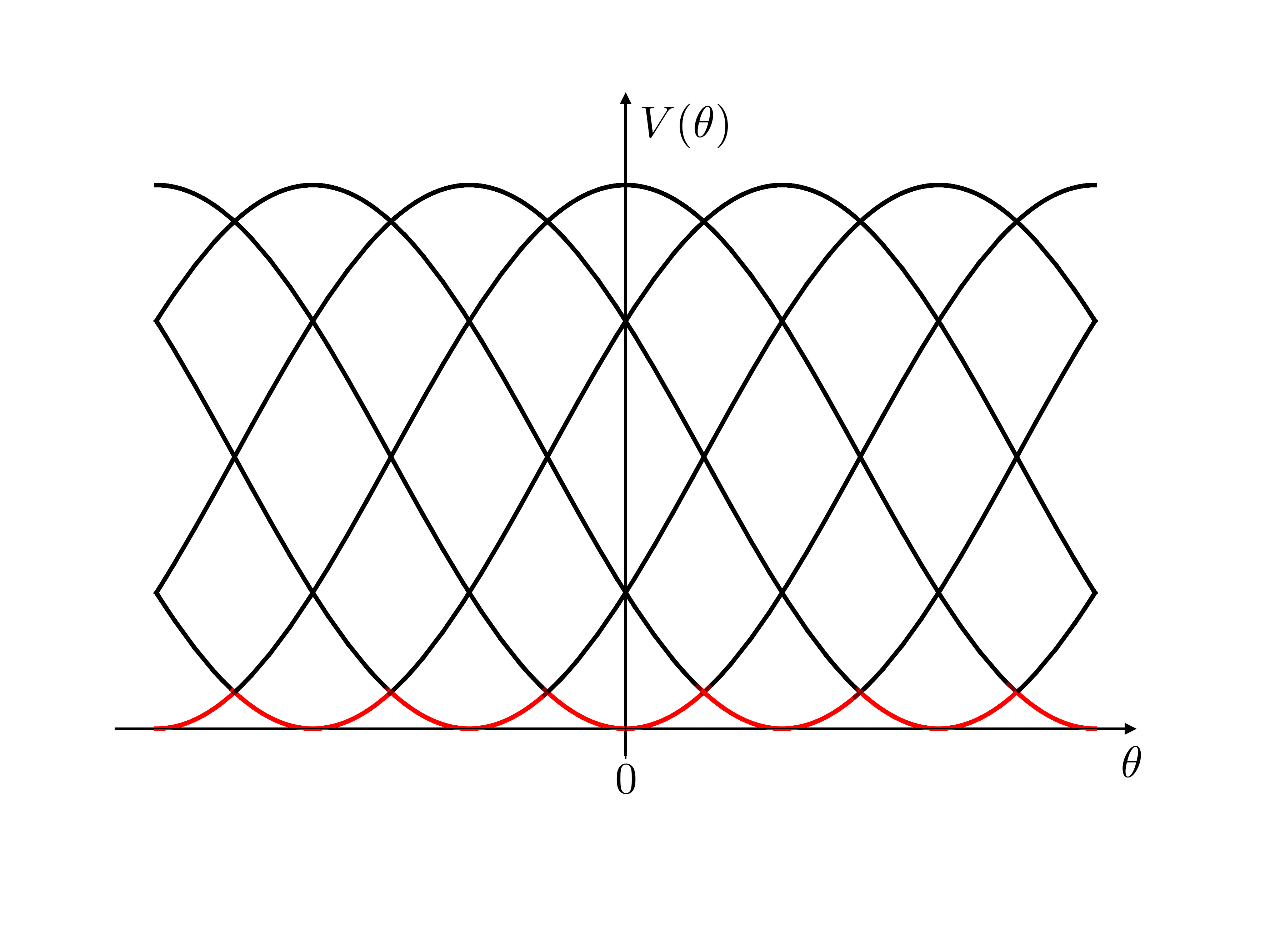}
\caption{Sketch of branches of $V(\theta)$ for pure $SU(6)$ YM theory according to a large-$N_c$ analysis. The red
portions of the curves give the true stable ground state energy density as a function of the
parameter $\theta$.  A refined analysis is necessary to determine if the black sections of the curves include metastable states at finite $N_c$, and
if so how many at each value of $\theta$.\label{fig:SU6Branches}} 
\end{figure}

Why do these pure glue metastable states exist, and what characterises them?  Roughly speaking the argument is that the vacuum energy density
$V(\theta)$ must be,  for all $N_c$,  both a $2\pi$ periodic function of the topological parameter $\theta$, $V(\theta)=V(\theta+2\pi)$, \emph{and} in
the large-$N_c$ limit have the form $V(\theta) = N_c^2 f(\theta/N_c)$ with $f(x)$ a $2\pi$-periodic function, and no other factors of $N_c$ 
appearing \cite{Witten:1998uka}.  (We emphasise
that here it is being assumed that there is no light QCD axion in the spectrum to relax the
CP-violating $\theta$-term.)  But these demands
seem to be contradictory as the second condition says the period of $V$ should be $2\pi N_c$.  The resolution, similarly to the case of the chiral Lagrangian,
is that $V(\theta)$ must be a \emph{multi-branched function}.  The true ground state energy density is then given by
\beq
V(\theta) =  {\rm min}_n  V^{(n)}(\theta)  \quad {\rm with} \quad V^{(n)}(\theta) = N_c^2 f\left(\frac{\theta + 2\pi n}{N_c}\right) ,
\eeq
where $n=0,\ldots,N_c-1$.  Namely each branch has periodicity $2\pi N_c$, but because of level crossings the true ground state energy density is correctly $2\pi$-periodic.  This multi-branch structure leads at every value of $\theta$ to a total of $N_c$ potential vacua, including at $\theta=0$, as illustrated schematically in
figure~\ref{fig:SU6Branches} for $SU(6)$.  In fact Witten \cite{Witten:1998uka} argued that in the formal $N_c\rightarrow \infty$ limit the function
$f(x) = x^2$ and that all states were local minima.  However at finite $N_c$ as the various branches
evolve and cross the character of the local critical point can change from absolutely stable, to metastable, to saddle point to local maximum.  So to really
enumerate the number of locally stable states in a finite $N_c$ theory at a given $\theta$ one needs a finer classification. 

Before turning to this it is worth mentioning that there is a simple expression for the gap in energy density at $\theta=0$
between the low-lying states, $n\ll N_c$, in the $N_c\rightarrow \infty$ limit:
\beq
 V^{(n)}(0) - V^{(0)}(0) \simeq \frac{(2\pi n)^2}{2} \chi
 \eeq
 where $\chi$ is the so-called topological susceptibility defined by
 \beq
 \chi = \int d^4x \langle Q(x)Q(0) \rangle \quad {\rm where} \quad Q = \frac{1}{8\pi^2}\Tr(G^{\mu\nu} {\tilde G}_{\mu\nu})~.
 \eeq
 Moreover, in the large-$N_c$ limit an order parameter distinguishing the $N_c$ different states at a given value of $\theta$ is~\cite{Gabadadze:2002ff}
 \beq
 \langle Q \rangle_n = (\theta +2\pi n) \chi~.
 \eeq
 (There are other possible order parameters, for example the QCD string tension in each $n$-state.)  Related to this, the $n\neq 0$ states are
 not invariant under $CP$ transformations even if $\theta = 0,\pi$, the two $CP$-invariant values of the topological angle. For the SM the topological
 susceptibility has a value that depends on the $\eta'$ mass as $\chi \sim ( f_{\eta'} m_{\eta'} )^2/6$, at least in the large-$N_c$ limit \cite{Witten:1979vv,Veneziano:1979ec}, while for pure $SU(3)$ gauge theory lattice simulations indicate $\chi \sim (190\MeV)^4$  \cite{Del_Debbio_2005} (though this
depends somewhat on how the lattice scale is set).
 
 Importantly in the large-$N_c$ limit an $\OO(N_c)$ subset of these states become highly metastable.  They are both locally stable, and have
 false vacuum tunnelling rates to lower states which have been argued, using an approach based on softly-broken supersymmetric QCD, to behave as $\Gamma_n /{\rm Volume} \sim \exp( - b N_c^4 )$ for some constant $b>0$  and so become absolutely stable in the formal $N_c\rightarrow \infty$ limit \cite{Shifman:1998if}. (We caution
 the reader that the computation of these decay rates in the far-from-supersymmetric limit is not under good control, so numerical factors and, in our opinion,
 even the $N_c^4$ scaling are not reliably established.\footnote{See also the discussion of Ref. \cite{Dine:2016sgq}.}  In addition there has been only a very limited study
 of this large-$N_c$ physics using lattice techniques \cite{DelDebbio:2006yuf,Bonati:2016tvi}.)

In recent work \cite{Aitken:2018mbb} arguments have been presented that, depending on the value of ${\theta}$, the number of locally stable vacua, $N_s$ (excluding the true ground state), is by a particular semiclassical analysis given by
\[
N_s =
\begin{cases}
2\left[\frac{N_c}{4}\right] \quad & {\theta}  = 0 \nonumber \\
\left[\frac{N_c}{4}\right] + \left[\frac{N_c+3}{4}\right] - 1 \quad & 0  < {\theta} < \pi/2 \nonumber \\
\left[\frac{N_c+1}{2}\right] -1 \quad & {\theta}  = \pi/2 \nonumber \\
\left[\frac{N_c+1}{4}\right] + \left[\frac{N_c+2}{4}\right] -1 \quad & \pi/2  < {\theta} < \pi \nonumber \\
2\left[\frac{N_c+2}{4}\right] -1\quad & {\theta}  = \pi \nonumber \\
\end{cases}
\]
where $\left[ x\right]$ denotes the largest integer $\leq x$.  So if we take this analysis as a guide
then, naively, at $N_c=3$ there are predicted to be, respectively, $(0,0,1,1,1)$ metastable states as ${\theta}$ varies in the given ranges, while
for $N_c=4$ there would be $(2,1,1,1,1)$ metastable states.\footnote{In Ref.~\cite{Halperin:1998rc} the large-$N_c$ effective 
Lagrangian was modified by including $1/N_c$ effects, possibly enabling an improved discussion of the metastable states of $SU(N_c)$ at finite $N_c$.  A richer
structure of metastable states was found in this analysis, with new metastable states being argued to exist both for pure $SU(3)_c$ Yang-Mills theory, and for $SU(3)_c$
coupled to $N$ light quarks. If correct this would strengthen our results.  We emphasise that no calculations that are fully under control have yet been performed in the 
interesting region of small $N_c$.}
It is particularly interesting that in this case there are metastable states
predicted at $\theta=0$.   

However, at small $N_c$, all the semiclassical arguments that we are aware of break down, and it is not clear from the analysis that has so far been performed
if there are metastable states at low $N_c$, in particular for $SU(3)_c$.   There are (yet) no reliable lattice studies of this question, so strictly speaking the situation regarding pure $SU(3)_c$ YM theory is unknown.  

\subsection{Possible relation to the hierarchy problem}

If we boldly assume that there are metastable states in pure $SU(3)_c$ at $\theta=0$ split from the true ground state by an energy density $\Delta V \sim \Lambda_{qcd}^4$
the refined Swampland de Sitter Conjecture, together with the results on quintessence in section~\ref{quint} (which equally apply to this case), then plausibly tells us that the SM with electroweak symmetry breaking scale $v\gg 50\TeV$ is in the swampland.  In other words, \emph{there are no solutions of the full UV gravitational theory (including 
all possible consistent compactifications if it is higher dimensional) that lead to the SM with an electroweak vev} $v\gg 50\TeV$.
This would be a partial resolution to the hierarchy problem and shares some features with previous attempts to link the hierarchy problem
with the swampland program \cite{Cheung:2014vva,Craig:2019fdy}.  In particular there is a failure of effective field theory reasoning, in that apparently
innocuous parameter regions of otherwise consistent quantum field theories are inconsistent when coupled to gravity.\footnote{One of the most studied
of the Swampland constraints is the Weak Gravity Conjecture which, among other things, states that QED with a single Dirac fermion of mass $m$ and gauge charge
$e$ is inconsistent with gravity if $|e|\leq m/\sqrt{2}\mpl$ \cite{ArkaniHamed:2006dz}.  This and the magnetic form of the conjecture have been suggested to possibly
provide an explanation of the weak-to-Planck scale hierarchy in extensions of the SM \cite{Cheung:2014vva,Craig:2019fdy}.}

Of course there are caveats to this statement, even if the SdSC is accepted as fact, and pure $SU(3)_c$ at $\theta=0$ has metastable states.  First, we have taken the
Yukawa couplings to be fixed as $v\rightarrow\infty$, but if, instead, these couplings were reduced so as to keep the ``hard'' quark masses coming from EWSB fixed at their
observed values, then the QCD sector of the theory would not possess metastable states, and there apparently would be no constraint on $v$.  We do not
have a strong argument against this reasoning.  However, we note that the limit where all the quark Yukawa couplings vanish is forbidden by the claimed absence of exact global
symmetries in theories coupled to gravity, which is one of the best supported of all Swampland Program conjectures.   In fact the quantitative Swampland Global Symmetry Conjecture \cite{Fichet:2019ugl,Daus:2020vtf} mentioned in the Introduction declares that there is a bound on how small more than one of the Yukawa couplings can simultaneously be, $y\gsim \exp(-\mpl^2/\Lambda^2)$, so if the cutoff $\Lambda$ of the effective 4d quantum field theory is close to the Planck scale the option of scaling all the Yukawa couplings to close to zero is forbidden.  This illustrates how the web of swampland conjectures might intertwine to limit the allowed SM parameter values.   

Second the arguments above have not constrained the case where the Higgs mass-squared parameter is large and positive so there is no traditional 
electroweak symmetry breaking.  This is the subject of a companion paper.

Third, we presently have poor understanding of the non-perturbative physics of QCD-like theories, especially as regards ``exotic" phenomena such as
metastable states.  It is logically possible that there are different features of full non-perturbative QCD (and the SM!) that are constrained by present, or future,
swampland program conjectures/results, and that these constraints are more powerful than the SdSC in limiting the available SM parameter values.

\acknowledgments

JMR and RPB gratefully thank the CERN Theory Group for hospitality during part of this work.  We thank Mike Teper for discussions of metastable
states in pure $SU(N_c)$ Yang-Mills theories, Arthur Hebecker for useful comments, and, especially, Matthew McCullough for extensive discussions.  RPB
is supported by a joint Clarendon and Foley-Bejar Scholarship from the University of Oxford and Balliol college.

\bibliography{QCDswampland}

\end{document}